\documentclass[hyper]{JHEP} % 10pt is ignored!

\usepackage{amssymb}

\usepackage{epsfig}
\newcommand\fverb{\setbox\pippobox=\hbox\bgroup\verb}

\newcommand\fverbdo{\egroup\medskip\noindent%

            \fbox{\unhbox\pippobox}\ }

\newcommand\fverbit{\egroup\item[\fbox{\unhbox\pippobox}]}

\newbox\pippobox

\title{ Mass Deformed L-BLG Theory From ABJ Theory}
\author{Tanay K. Dey\\
Institute of Physics\footnote{Short term visitor},\\
Bhubaneswar 751 005, INDIA.\\
E-mail: \email{tanay.dey@gmail.com}}

\author{Kamal L. Panigrahi\\
Department of Physics, \\
Indian Institute of Technology Ropar, Rupnagar-140001, India\\
and \\
Department of Physics \& Meteorology, \\
Indian Institute of Technology Kharagpur,
Kharagpur, 721 302, INDIA.\\
E-mail: \email{panigrahi@iitrpr.ac.in,panigrahi@phy.iitkgp.ernet.in}}

\abstract{We construct mass deformed $SU(N)$ L-BLG theory together with $U(M-N)_k$ Chern-Simons theory. This mass deformed
L-BLG theory is a low energy world volume theory of a stack of $N$ number of M2-brane far away from $\bf{C^4/Z}_k$
singularity. We carry out this by defining a special scaling limit of the fields of this theory and simultaneously sending
the Chern-Simons level to infinity.}


\def\tr{\mathrm{Tr}}

\def\tA{\tilde{A}}

\newcommand{\tX}{\tilde{X}}

\newcommand{\tB}{\tilde{B}}

\newcommand{\tD}{\tilde{D}}

\newcommand{\mL}{\mathcal{L}}

\def \tY{\tilde{Y}}
\def\pb #1{\left\{#1\right\}}

\begin{document}
%%%%%%%%%%%%%%%%%%%%%
%%%%Introduction %%%%%%%%%
%%%%%%%%%%%%%%%%%%%%
\section{Introduction}\label{first}
Since the last few years there has been a huge interest in constructing three
dimensional low energy world volume theories of coincident
M2-branes with the aim of finding out the dual boundary gauge theory of a 11-dimensional gravity theory with
geometric structure $AdS_4 \times S^7$. With this aim in view Bagger, Lambert \cite{Bagger1,Bagger2,Bagger3} and
Gustavsson \cite{Gustav,Gustav1} (BLG) first
constructed a Chern-Simons matter theory with ${\cal N}=8$ supersymmetry and $SO(8)$ global symmetry based on $3$-algebra.
The gauge group of this theory is restricted to $SO(4)$. By complexifying the matter fields, the BLG theory
can be rewritten as a Chern-Simons gauge theory with the gauge group $SU(2)\times SU(2)$ generated by ordinary Lie algebra.
One of the gauge groups is associated with Chern-Simons level, $k$ and the other, with $-k$\cite{Raam}.
Immediately, it was discovered that the moduli space of BLG theory describes world volume theory of only two coincident
M2-branes\cite{Raam,Bandres1,Papa,Gaunt,Gomis,Paul,Paul1}. In order to go beyond this two brane theories, it was suggested
in \cite{Gomis,Sergio,Ming,Bandres2,Gomis1,Bobby,Herman} to use a $3$-algebra with a metric having indefinite Lorentzian signature.
The resulting Lorentzian-BLG (L-BLG) theory still preserves ${\cal N}=8$ supersymmetry at the classical level, though its
quantum version lacks proper interpretation.\\

\noindent
Inspired by BLG theory and subsequent developments, Aharony, Bergman, Jafferis and Maldacena (ABJM)
constructed the low energy effective theory for $N$ number of multiple M2-branes sitting at the singularity of the $\mathbf{C^4/Z}_k$
orbifold \cite{Abjm}. Instead of the $3$-algebra, this construction is fully based on an ordinary Lie algebra. The gauge group of this theory is
$U(N)_k\times U(N)_{-k}$ with ${\cal N}=6$ explicit supersymmetry and $SU(4)_R\times U(1)$ global symmetries\cite{Benna}.
It has two parameters, $N$ and $k$ which take only integer values. The field content of the theory are $N\times N$ matrices
transforming in the bi-fundamental representation of the gauge group. The dual gravity theory is the M theory living on
$AdS_4 \times S^7/{\bf Z}_k$.  Although it seems to be different from that of the original L-BLG theory, this gauge theory
also admits a $3$-algebra interpretation \cite{Bagger,Martin}.\\

\noindent
In \cite{Honma}(also in \cite{Honma1}), a suggestion was put forword to produce the L-BLG theory from ABJM theory. According to
their proposal, L-BLG theory can be obtained from the ABJM theory by sending the Chern-Simons level, $k$ to
infinity and set some of the fields of the theory to zero at the same time so that they decouple. In \cite{Anto}, the
suggestion was more refined  and it was pointed out that the scaling limit alone
is not sufficient to produce the L-BLG theory correctly from
ABJM theory. One needs  to associate an extra ghost multiplet also with ABJM
theory which is decoupled from the theory itself but effectively couples with it on redefinition of the fields. With the same spirit in \cite{Honma2}, the extended L-BLG theory with two pairs of Lorentzian generators is derived by taking a scaling limit of several quiver Chern-Simons theories obtained from different orbifoldings of the ABJM action.\\

\noindent
Shortly after ABJM theory, as a further generalization, another model is proposed by Aharony, Bergman, Jafferis\cite{Abj})
named as the ABJ model. This model involves modification of ABJM gauge group $U(N)_k \times U(N)_{-k}$ to $U(M)_k \times U(N)_{-k}$
with $M \ge N$, in the
Chern-Simons matter fields kept unchanged . In the gravity picture, the brane construction is equivalent to the low energy
 $(M-N)$ fractional M2-branes sitting at the $\mathbf{C^4/ Z}_k$ singularity in one sector and $N$ M2-branes freely
moving around on the other. The geometric structure of the gravity
theory remains same as that of the ABJM theory but with an
additional torsion flux that takes values in $H^4(S^7/{\bf
Z}_k,{\bf Z}) = {\bf{ Z}}_k$.  Since ABJ theory is slightly more
general than ABJM theory it deserves a similar study. More
precisely, one should take the same scaling limit as taken in ABJM
theory and check whether this theory also reproduces L-BLG theory.
With this aim in view, a study is initiated in \cite{Kluson} with
a redefined scaling limit of fields of the theory. This study was
semi-complete, as it concentrated only in the massless sector.
However in order to go towards a non-relativistic limit of this
theory with some of the supersymmetries of the relativistic theory
still preserved, we do require a mass term. The non-relativistic
version of the 3-dimension L-BLG theory could be useful to study
the strongly coupled condensed matter system.  With this
motivation, in this paper we extend the study further with the
inclusion of mass term in the theory. Although our aim is to
produce the mass deformed L-BLG theory in this work, one can also
extend this study to construct the full non-relativistic version
of the theory which we leave for future.\\

\noindent
In this work, we begin with the Lagrangian of the ABJ theory and introduce a suitable supersymmetric mass
deformation as in \cite{Hosom,Hosom1,Verlinde} and then take the appropriate scaling limit on the fields and at the same time
send the Chern-Simons level, $k$ to infinity. In this process, we get a mass deformed $SU(N)$ L-BLG theory together with $U(M-N)_k$
Chern-Simons theory. This L-BLG theory is a low energy world volume theory of $N$ number of $M2$ branes stack far
away from ${\bf C^4/Z}_k$ singularity.\\

\noindent
The rest of this paper is organized as follows. In section-2, we briefly review the
main aspects of the ABJ theory. In section-3, we
introduce the maximally supersymmetric mass deformation term in to the theory.
Section-4 is devoted to the scaling limits on the fields of the theory and sending the level $k$ to infinity to finally arrive at the $U(N)$ mass deformed L-BLG theory together with  $U(M-N)_k$ Chern-Simons theory. In section-5, we present our conclusions.

%%%%%%%%%%%%%%%%%%%%%%%%%%%%%%%%%%5
\section{A Brief Summary of ABJ Theory}\label{second}
In this section we briefly recapitulate few basic facts about the ABJ theory. As mentioned in the introduction, the ABJ theory
is the generalized version of the ABJM theory. The gauge group of this theory is $U(M)_k \times U(N)_{-k}$ with $SU(4)$ global
symmetry. The quantum consistency demands  $|M-N|\geq k$ \cite{Abj}. Unlike the
ABJM theory, which has two parameters, ABJ theory contains
one extra parameter due to the addition of a fractional M2 brane. These parameters are $M,N$ and $k$  all of which take integer values.
The field content of this theory are:
\begin{itemize}
\item two gauge fields, $ A_\mu^{(L)}$ and $A_\mu^{(R)}$ associated with the groups $U(M)$
and $U(N)$ respectively,
\item four complex scalar fields, $Y^A (A=1,2,3,4)$ which are $M\times N$ matrix valued, together with their hermitian
conjugates, $Y^\dag_A$ that are $N\times M$ matrix valued,
\item $M\times N$ matrix valued fermions $\psi_A$ and their hermitian conjugates $\psi^{A\dag}$ given by $N\times M$ matrix.
\end{itemize}
Fields with upper indices, as specified above, transform in the $\mathbf{4}$ of $R$
symmetry $SU(4)$ group and those with lower indices transform in the $\overline{\mathbf{4}}$ representations. The Lagrangian
for this fields has the following form \cite{Benna}
\begin{eqnarray}\label{defABJL}
\mL &=&-\tr (D^\mu Y^\dag_A D_\mu Y^A)-i \tr (\overline{\psi}^{A}\gamma^\mu D_\mu\psi_A)+V_{pot}+\mL_{CS}\nonumber \\
&&-i\frac{2\pi}{k} \tr(Y_A^\dag Y^A\overline{\psi}^{B}\psi_B -Y^A Y^{\dag}_A\psi_B\overline{\psi}^{B} )
 +2i\frac{2\pi}{k} \tr (Y^\dag_A Y^B\overline{\psi}^{A}\psi_B -Y^A Y_B^\dag\psi_A\overline{\psi}^{B} )\nonumber \\
 &&+ i\frac{2\pi}{k} \epsilon_{ABCD} \tr (Y^A\overline{\psi}^{B}Y^C \overline{\psi}^{D})-i\frac{2\pi}{k}
 \epsilon^{ABCD}\tr(Y^\dag_A \psi_B Y^\dag_C \psi_D).
 \end{eqnarray}
 Where
 $\mL_{CS}$ given in (\ref{defABJL}) is a Chern-Simons term
and $V_{pot}$ is a sextic scalar potential and they have the following forms
\begin{eqnarray}{\label{CSact}}
\mL_{CS}&=&\frac{k}{4\pi}\epsilon^{\mu\nu\lambda}\tr[A^{(L)}_\mu\partial_\nu A_\lambda^{(L)}+
\frac{2i}{3}A_\mu^{(L)}A_\nu^{(L)}A_\lambda^{(L)}]\nonumber \\
&-&\frac{k}{4\pi} \epsilon^{\mu\nu\lambda}\tr[A_\mu^{(R)}\partial_\nu A_\lambda^{(R)}+\frac{2i}{3}A_\mu^{(R)}
A_\nu^{(R)}A_\lambda^{(R)}] \ ,\nonumber \\
V_{pot}&=& \frac{4\pi^2}{3k^2}\tr [Y^AY_A^\dag Y^B Y_B^\dag Y^C Y_C^\dag +Y_A^\dag Y^A Y_B^\dag Y^B Y_C^\dag Y^C\nonumber \\
&&+4 Y^AY_B^\dag Y^C Y_A^\dag Y^B Y_C^\dag -6 Y^A Y_B^\dag Y^B Y_A^\dag Y^C Y_C^\dag] \ .
\end{eqnarray}
Further,  the covariant derivatives for the scalars are defined as
\begin{eqnarray}{\label{covscalar}}
D_\mu Y^A=\partial_\mu Y^A +iA^{(L)}_\mu Y^A-iY^A A^{(R)}_\mu \ , \quad
D_\mu Y^\dag_A=\partial_\mu Y^\dag_A -iY^\dag_A A_\mu^{(L)}+i A_\mu^{(R)}Y^\dag_A \ ,\nonumber\\
\end{eqnarray}
and that for fermions are defined as
\begin{eqnarray}{\label{covfermi}}
D_\mu \psi_A=\partial_\mu \psi_A+i A^{(L)}_\mu \psi_A-i\psi_A A_\mu^{(R)}\ , \quad D_\mu \psi^{\dag A}=
\partial_\mu \psi^{A\dag}-i\psi^{\dag A}A_\mu^{(L)}+ iA_\mu^{(R)}\psi^\dag_A\ .\nonumber\\
\end{eqnarray}
The three dimensional gamma matrices are
$$ \gamma^0=i\sigma^2, \quad \gamma^1= \sigma^1, \quad \gamma^2=\sigma^3,$$ where the $\sigma$'s are the Pauli matrices.\\

\noindent
In \cite{Bagger}, the general form for the action for a three dimensional scale-invariant theory
with ${\cal N}=6$ supersymmetry, $SU(4)$ $R$ symmetry and $U(1)$ global symmetry was found by starting with a $3$-algebra in
which the triple product is not antisymmetric. The field content is the same as described above with the matter
fields now taking values in the $3$-algebra:
\begin{equation}
Y^A=T^\alpha Y_\alpha^A \, \psi_A=T^\alpha\psi_{A\alpha} \, A_\mu=A_{\mu \alpha\beta}T^\alpha\otimes T^\beta.
\end{equation}
In what follows, we use the notation as in \cite{Bandres}. The complex scalar $Y^A$ have explicit form
\begin{eqnarray}
(Y^A)^\alpha_{\ \hat\alpha} \ \in (\mathbf{M,\overline{N}}) \  , \quad (Y^\dag_A)^{\hat\alpha}_{\ \alpha} \in (\mathbf{\overline{N},M}) \ ,
\end{eqnarray}
where $\alpha=1,\cdots, M \ , \hat\alpha=1,\cdots , N$. The gauge field  $(A^{(L)}_\mu)^\alpha_{\ \beta}$  and
$(A^{(R)}_\mu)^{\hat\alpha}_{\ \hat\beta}$ are hermitian matrices of  $U(M)$ and $U(N)$ respectively. Then clearly
\begin{eqnarray}
(D_\mu Y^A)^\alpha_{\ \hat\alpha}=\partial_\mu (Y^A)^\alpha_{\  \hat\alpha}+i(A^{(L)}_\mu)^\alpha_{\ \beta}(Y^A)^{\beta}_{\
\hat\alpha}- i(Y^A)^\alpha_{\ \hat\beta}(A^{(R)}_\mu)^{\hat\beta}_{ \ \hat\alpha}\nonumber \\
(D_\mu Y^\dag_A)^{\hat\alpha}_{\ \alpha}=\partial_\mu (Y^\dag_A)^{\hat\alpha}_{ \ \alpha}-i(Y^\dag_A)^{\hat\alpha}_{\
\beta}(A^{(L)}_\mu)^\beta_{\ \alpha}+i(A_\mu^{(R)})^{\hat\alpha}_{\ \hat\beta}(Y^\dag_A)^{\hat\beta}_{\ \alpha}.
\end{eqnarray}
So far we have not introduced the mass term in the theory. However, as  mentioned in the introduction, one of the way
to get to the  non-relativistic limit of ABJ theory is the introduction of suitable mass term which preserves some of
the supersymmetries of the original relativistic version. Though we are not going to study the non-relativistic theory in this paper,
but we will introduce it in the next section to produce the mass deformed L-BLG theory from the mass deformed ABJ theory.
This construction can be useful to construct the non-relativistic version of the theory which we leave for future work.

%%%%%%%%%%%%%%%%%%%%%%%%%%%%%%%%%%%%%%%%%%%55

\section{Mass Deformed ABJ Theory}{\label{nrabj}}
In order to introduce the mass term, we focus on maximally supersymmetric mass deformation. It turns out that the suitable form of
this mass term is\cite{Hosom,Hosom1,Verlinde,Kwon}
\begin{eqnarray}
V_{mass}&=& \tr (M^C_AM^B_CY^\dag_B Y^A)+i \tr(M^B_A\overline\psi^{A}\psi_B)-\frac{4\pi}{k}\tr (M^C_BY^AY_A^\dag Y^BY_C^\dag)\nonumber\\
 &=& m^2 \tr(Y^\dag_A Y^A)+ im \tr(\overline\psi^{a} \psi_a)-im \tr(\overline\psi^{a'}\psi_{a'})\nonumber \\
&&-\frac{4\pi}{k}m\tr(Y^aY^{\dag}_{[a}Y^bY_{b]}-Y^{a'}Y^\dag_{[a'}Y^{b'}Y_{b']}^\dag).
\end{eqnarray}
Here $m$ is the mass parameter and the diagonal form of mass matrix $M^A_B = m\, {\rm diag}(1,1,-1,-1)$. The matrix $M^B_A$ satisfy the following constraints
\begin{eqnarray}
M^\dag = M, \quad\quad \tr M = 0 \quad  {\rm and} \quad M^2= m^2 {\mathbf{1}}\nonumber\\
M^A_B Y^B = m Y^a -m Y^{a'}, \quad M^B_A Y^\dag_B = m Y^\dag_a - m Y^\dag_{a'}\nonumber\\
M^A_B \psi_A = m \psi^b -m \psi^{b'}, \quad M^B_A \overline\psi^{A} = m \overline\psi^{b} - m \overline\psi^{b'}
\end{eqnarray}
The introduction of mass term breaks explicitly the original $SU(4)\times U(1)$ R-symmetry and down
to $SU(2)_L \times SU(2)_R \times U(1) $\cite{Hosom,Hosom1,Verlinde}. We set $A=(a,a')$, where $a$
and $a'$ are two $SU(2)$ indices take values 1, 2 and 3, 4 respectively. We used the following convention: $$X_{[a}X_{b]}=X_aX_b-X_bX_a.$$
In other words we can say that the mass term of the ABJ theory can be introduced in the same way as in the ordinary ABJM theory.
On the other hand, our realization is that the gauge group is $U(M)\times U(N)$ makes impact on the structure of vacuum manifold.\\

%%%%%%%%%%%%%%%%%%%%%%%%%%%%%%%%%%%%%%%%%%%%%%%%%%%%%%%
\section{Scaling Limit of Mass Deformed ABJ Theory}
In this section we construct the mass deformed L-BLG theory from the mass deformed ABJ theory following the suggestions given
in \cite{Honma,Anto}. Basically we take scaling limit of the fields of the ABJM theory. Therefore as the first step we presume
that $A_\mu^{(L)}$ takes the following form
\begin{eqnarray}
A_\mu^{(L)}=\left(\begin{array}{cc}
A_{11 \mu}^{(L)} & A_{12\mu}^{(L)} \\
A_{21 \mu}^{(L)} & A_{22\mu}^{(L)} \\
\end{array}\right) \ .
\end{eqnarray}
Here $A_{11\mu}^{(L)}$ is $(M-N)\times (M-N)$ matrix.  $A_{12\mu}^{(L)}$ and $A_{21\mu}^{(L)}$ are $(M-N)\times N$ and
$N\times (M-N)$ matrices respectively. $A_{22\mu}^{(L)}$ is $N\times N$ square matrix. We further introduce the field,
$B_\mu$ defined as
\begin{equation}
A_{22\mu}^{(L)}=A_\mu- \frac{1}{2}B_\mu\ , \quad A_{\mu}^{(R)}= A_\mu+\frac{1}{2}B_\mu \ .
\end{equation}
The significance of this notation is that the  $B_\mu$ is an auxiliary
field. To see this note that Chern-Simons term of (\ref{CSact}) implies
\begin{eqnarray}
&&\frac{k}{4\pi}\int d^3x \epsilon^{\mu\nu\lambda}[\tr(A^{(L)}_\mu \partial_\nu A^{(L)}_\lambda)
-\tr (A^{(R)}_\mu \partial_\nu A^{(R)}_\lambda)]\nonumber \\
&=&\frac{k}{4\pi} \int d^3x\epsilon^{\mu\nu\lambda} [\tr(A_{11\mu}^{(L)}\partial_\nu A_{11\lambda}^{(L)})+ 2\tr
(A_{12\mu}^{(L)}\partial_\nu A_{21\lambda}^{(L)})- \tr (B_\mu (\partial_\nu A_\lambda-\partial_\lambda A_\nu) )]\nonumber\\
\end{eqnarray}

and
\begin{eqnarray}
&&\frac{k}{4\pi}\frac{2i}{3} \int d^3x\epsilon^{\mu\nu\lambda} [\tr(A_\mu^{(L)}A_{\nu}^{(L)}A_{\lambda}^{(L)}-
A_\mu^{(R)}A_\nu^{(R)}A_\lambda^{(R)})]\nonumber \\
&=&\frac{k}{4\pi} \frac{2i}{3} \int d^3x\epsilon^{\mu\nu\lambda} \tr[A^{(L)}_{11\mu}A^{(L)}_{11\nu}A^{(L)}_{11\lambda}
+3A_{11\mu}^{(L)}A_{12\nu}^{(L)}A_{21\lambda}^{(L)}\nonumber\\
&&+3A_{22\mu}^{(L)}A_{21\nu}^{(L)}A_{12\lambda}^{(L)}+ A_{22\mu}^{(L)}A_{22\nu}^{(L)}A_{22\lambda}^{(L)}
-A_\mu^{(R)}A_\nu^{(R)}A_\lambda^{(R)} ]
\end{eqnarray}

and together we obtain

\begin{eqnarray}\label{csl}
\mL_{CS}&=&\frac{k}{4\pi}\epsilon^{\mu\nu\lambda}\tr [A_{11\mu}^{(L)}\partial_\nu A^{(L)}_{11\lambda} +\frac{2i}{3}
A_{11\mu}^{(L)}A_{11\nu}^{(L)}A_{11\lambda}^{(L)}\nonumber\\
&+&2A_{12\mu}^{(L)}\partial_\nu A_{21\lambda}^{(L)} + 2i (A_{11\mu}^{(L)}A_{12\nu}^{(L)} A_{21\lambda}^{(L)}+
A_{22\mu}^{(L)}A_{21\nu}^{(L)}A_{12\lambda}^{(L)})\nonumber \\
&-&B_\mu(\partial_\nu A_\lambda-\partial_\lambda A_\nu+i[A_\nu,A_\lambda]) -\frac{i}{6}B_\mu B_\nu B_\lambda].
\end{eqnarray}
Since in the above form there is no kinetic term for $B_\mu$ field, it confirms the claim that $B_\mu$ is an auxiliary field.
 Note that the $N\times N$ matrix gauge field $A_{22\mu}$ is not decoupled from the massive  $N\times (M-N)$ matrix gauge
field $A_{21\mu}$ which is much needed to construct the mass
deformed L-BLG theory on $N$ number of M2-branes.  To decouple
this we consider the scaling limit as in undeformed ABJ
theory\cite{Kluson}. The form of the scaling limit is
\begin{eqnarray}\label{scalgauge}
A_{11\mu}^{(L)}&=&A_{11\mu}^{(L)}\ , \quad A_{12\mu}^{(L)}=\epsilon^2 \tA_{12\mu}
\ , \quad  A_{21\mu}^{(L)}= \epsilon^2\tA_{21\mu}\ , \nonumber \\
A^{(L)}_{22\mu}&=& A_\mu-\frac{1}{2}\epsilon B_\mu \ ,\quad A^{(R)}_\mu= A_\mu+\frac{1}{2}\epsilon B_\mu,
\end{eqnarray}
where $\epsilon$ is the small parameter which controls the scaling limit and finally we  take $\epsilon\rightarrow 0$.
Note that $A_\mu$ and $B_\mu$ belong to the algebra of $\mathbf{U}(N)$.\\

\noindent
We then plug in redefined gauge fields of (\ref{scalgauge}) into the Chern-Simons Lagrangian of (\ref{csl}) and  we get
\begin{eqnarray}
\mL_{CS}&=&\frac{k}{4\pi}\epsilon^{\mu\nu\lambda} \tr(A_{11\mu}^{(L)}\partial_\nu
A_{11\lambda}^{(L)}+\frac{2i}{3}A^{(L)}_{11\mu}A^{(L)}_{11\nu}A^{(L)}_{11\lambda}\nonumber \\
&+& 2 \epsilon^4 \tilde A_{12\mu}\partial_\nu \tilde A_{21\lambda} + 2i\epsilon^4 A_{11\mu}^{(L)}
\tilde A_{12\nu}\tilde A_{21\lambda}+2i \epsilon^4 A_{22\mu}^{(L)}\tilde A_{21\nu}\tilde A_{12\lambda})\nonumber \\
&+&\frac{k}{4\pi} \epsilon^{\mu\nu\lambda}\Big[-\epsilon B_\mu (\partial_\nu A_\lambda-\partial_\lambda A_\mu+i[A_\nu, A_\lambda])
+{\cal O}(\epsilon^3)\Big]\nonumber \\
&\equiv& \mL^{(1)}+\mL^{(2)},
\end{eqnarray}
where
\begin{eqnarray}
\mL^{(1)}&=&\frac{k}{4\pi}\epsilon^{\mu\nu\lambda} \tr(A_{11\mu}^{(L)}\partial_\nu
A_{11\rho}^{(L)}+\frac{2i}{3}A^{(L)}_{11\mu}A^{(L)}_{11\nu}A^{(L)}_{11\lambda})\nonumber \\
&+& 2 \epsilon^4 \tilde A_{12\mu}\partial_\nu \tilde A_{21\lambda} + 2i\epsilon^4 A_{11\mu}^{(L)}
\tilde A_{12\nu}\tilde A_{21\lambda}+2i \epsilon^4 A_{22\mu}^{(L)}\tilde A_{21\nu}\tilde A_{12\lambda})
\end{eqnarray}
and
\begin{eqnarray}
\mL^{(2)}&=&-\frac{k\epsilon}{4\pi}\epsilon^{\mu\nu\lambda} \Big[ \tr B_\mu(\partial_\nu A_\rho-
\partial_\lambda A_\mu+i [A_\nu, A_\lambda]) - {\cal O}(\epsilon^2)\Big]
\end{eqnarray}
Notice that in order to decouple the massive states we should keep $k$ unscaled in the first part of the
Lagrangian, $\mL^{(1)}$. However, in the second part we should scale $k=\frac{1}{\epsilon}\tilde{k}$ and keep
$\tilde k$ finite in the limit $\epsilon \rightarrow 0$ and sending $k$ at infinity. Finally
in the limit $\epsilon \rightarrow 0$ we end up with the following form
\begin{eqnarray}
\mL^{(1)}&=&\frac{k}{4\pi}\epsilon^{\mu\nu\lambda} \tr(A_{11\mu}^{(L)}\partial_\nu
A_{11\lambda}^{(L)}+\frac{2i}{3}A^{(L)}_{11\mu}A^{(L)}_{11\nu}A^{(L)}_{11\lambda}) \ ,\nonumber\\
\mL^{(2)}&=&- \frac{\tilde{k}}{4\pi}\epsilon^{\mu\nu\lambda} \tr B_\mu F_{\nu\lambda} \ , \quad
F_{\nu\lambda}=\partial_\nu A_\lambda-\partial_\lambda A_\mu+i [A_\nu, A_\lambda].
\end{eqnarray}
We now try to find out the physical interpretation of the above results. The gauge fields of the Lagrangian density,
$\mL^{(1)}$ are $(M-N)\times (M-N)$ matrices. Therefore, this Lagrangian should describe $U(M-N)_k$ Chern-Simons
theory living on the world-volume of fractional M2-branes that are localized at the origin of $\mathbf{C^4/Z}_k$.
 On the other hand the fields of the Lagrangian density, $\mL^{(2)}$ are $N\times N$ matrices. Thus we have $U(N)$
Chern-Simons gauge theory for the $N$ number of M2-brane stack far away from the $\mathbf{C^4/Z}_k$. For further
analysis of this theory we split $B_\mu $ and $A_\mu$ gauge fields into $U(1)$ and
$SU(N)$ parts as
\begin{eqnarray}\label{SU}
A_\mu=A_\mu^0\mathrm{I}_{N\times N}+\tA_\mu \ , \quad   \tr \tA_\mu=0 \ , \nonumber \\
B_\mu= \epsilon^2 B_\mu^0\mathrm{I}_{N\times N}+ \tB_\mu \ , \quad  \tr\tB_\mu=0,
\end{eqnarray}
and then rescale the $U(1)$ part of $B$ field with $\epsilon^2$. $\mL^{(2)}$ then looks like
\begin{equation}
\mL^{(2)}=-\frac{\epsilon^2\tilde{k} N}{4\pi}\epsilon^{\mu\nu\lambda} B^0_\mu F^0_{\nu\lambda} - \frac{\tilde{k}}{4\pi}
\epsilon^{\mu\nu\lambda}\tr\tB_\mu \tilde{F}_{\nu\lambda}= -\frac{\tilde{k}}{4\pi}\epsilon^{\mu\nu\lambda}\tr\tB_\mu
\tilde{F}_{\nu\lambda} \ .
\end{equation}
Which is precisely the gauge part of L-BLG theory for the identification of $\tilde{k}=\pi$.\\

\noindent
Inspired by the above result, we now consider the scaling limit of matter fields $Y^A,\psi_A$ and scale them in the following way
\begin{eqnarray}\label{scalscalar}
Y^A=\left(\begin{array}{cc}
\epsilon^2 Z^A \\ \frac{1}{\epsilon}Y^A_+ \mathrm{I}_{N\times N}+\tY^A \\ \end{array}\right) \ , \quad \tr \tY^A=0, \nonumber \\
\psi_A= \left(\begin{array}{cc}\epsilon^2 \chi_A \\ \frac{1}{\epsilon}\psi_{+A}
\mathrm{I}_{N\times N}+\theta_A \end{array}\right) \ ,\quad \tr \theta_A=0,
\end{eqnarray}
where $Z^A$, $\chi_A$ are $M-N\times N$ matrices and $\tY^A$, $\theta_A$ are $N\times N$ matrices. For analysing the kinetic
term of the scalar field we first compute $D_\mu Y^A$ using the definition
of the covariant derivative of (\ref{covscalar}).
\begin{eqnarray}\label{dmu1}
D_\mu Y^A &=&\left(\begin{array}{cc} \epsilon^2 \partial_\mu Z^A+ i\epsilon^2 A_{11\mu}^{(L)}Z^A- Z
i\epsilon^2 A^{(R)}_\mu+ i \epsilon^2 \tilde A_{12\mu}(\frac{1}{\epsilon}Y_+^A \mathrm{I}_{N\times N}+\tY^A) \\
\frac{1}{\epsilon}\partial_\mu Y_+^A \mathrm{I}_{N\times N}+ \partial_\mu \tY^A+
i\left[A_\mu,\tY^A\right]- i Y_+^A
B_\mu -\frac{i}{2}\epsilon\pb{B_\mu,\tY^A} \\ \end{array}\right) \nonumber \\
&\Rightarrow&\left(\begin{array}{cc}  O(\epsilon)_{(M-N)\times N}\\
\frac{1}{\epsilon}\partial_\mu Y_+^A \mathrm{I}_{N\times N}+
\partial_\mu \tY^A+ i\left[A_\mu,\tY^A\right]- iY_+^A(\epsilon^2 B^0_\mu \mathrm{I}_{N\times N}+ \tB_\mu)
-\frac{i}{2}\epsilon\pb{B_\mu ,\tY^A} \\ \end{array}\right)\nonumber \\
& \equiv &\left(\begin{array}{cc} O(\epsilon)_{(M-N)\times N} \\
\frac{1}{\epsilon}\partial_\mu Y_+^A \mathrm{I}_{N\times N}+ \tD_\mu \tY^A-
\frac{i}{2}\epsilon\pb{\tB_\mu,\tY^A} \\ \end{array}\right).
\end{eqnarray}
Here we have defined
\begin{equation}
\tD_\mu \tY^A=\partial_\mu \tY^A+i[A_\mu,\tY^A]-iY_+^A \tB_\mu \ .
\end{equation}
In the similar way we find that
\begin{eqnarray}\label{dmu2}
D_\mu Y^\dag_A&=&
\left(\begin{array}{cc}
O(\epsilon)_{N\times (M-N)} &  \quad
 \frac{1}{\epsilon}
\partial_\mu Y_{+A}^\dag \mathrm{I}_{N\times N}+
(\tilde{D}_\mu \tY_A)^\dag
+
\frac{i\epsilon}{2}\pb{\tB_\mu,\tY_A^\dag}\end{array}
\right),
 \end{eqnarray}
where
\begin{equation}
(\tilde{D}_\mu Y_A)^\dag=\partial_\mu \tY_A^\dag-i[\tY_A^\dag,A_\mu] + iY_{A+}^\dag\tB_\mu \ .
\end{equation}
Using equation (\ref{dmu1}) and (\ref{dmu2}), we rewrite the kinetic term for $Y_A$ in the following form
\begin{eqnarray}\label{kintermsc}
\tr (D_\mu Y^\dag_A D^\mu Y^A )&=&\frac{N}{\epsilon^2} \partial_\mu Y_{+A}^\dag\partial^\mu Y^A_+ +
\tr (\tD_\mu \tY^\dag_A \tD^\mu \tY^A)\nonumber \\
&-&i\partial_\mu Y^\dag_{+A}\tr(\tB^{\mu}\tY^A) +i\partial_\mu Y^A_+\tr (\tB^{\mu}\tY_A^\dag).
\end{eqnarray}
Note that the first term diverges in the limit $\epsilon\rightarrow 0$. Therefore it seems that the scaling limit
is not complete. For this, following the suggestion given in \cite{Anto}, we add an
extra ghost term for the bosonic fields in the ABJ Lagrangian. This ghost term is a $U(1)$ multiplet containing four complex
bosonic fields $U^A$. The form of this ghost term is
\begin{equation}\label{addional}
\tr(\partial_\mu U_A^\dag \partial^\mu
U^A ) .
\end{equation}
Note that in (\ref{addional}) we have considered a ``wrong" sign
compared to the kinetic term since $U^A$ is a ghost field.
Following the argument as in \cite{Anto} for the ABJM theory, it
is natural to expect that the original ABJ action with similar
``ghost" term reduces to L-BLG action. Addition of the ghost,
results in an indefinite kinetic-term signature arising of a
manifestly definite ABJ action through regular scaling limit.
Although the extra ghost term is decoupled at the level of ABJ
theory, it is effectively coupled through the following
redefinition of the field:
\begin{equation}\label{redU}
U^A=-\frac{1}{\epsilon}Y_+^A {\rm I}_{N\times N}+
\epsilon\frac{1}{N}Y^A_-{\rm I}_{N\times N}\
\end{equation}
in the process when we implement the scaling limit. Note also that we have introduced a new scalar field, $Y^A_-$ through the
redefinition of the field. It will turn out that this new field plays a pivotal role in L-BLG theory. Finally, using (\ref{redU}),
the kinetic term (\ref{kintermsc}) together with the ghost term (\ref{addional}) reduces to a finite value in the limit of
$\epsilon\rightarrow 0$ and we get:
\begin{eqnarray}\label{kinfinal}
& & \tr(\partial_\mu U_A^\dag \partial^\mu U^A) -\frac{N}{\epsilon^2} \partial_\mu Y_{+A}^\dag
\partial^\mu Y^A_+-\tr (\tD_\mu \tY^\dag_A \tD^\mu \tY^A)\nonumber \\
&&+i\partial_\mu Y_{+A}^\dag\tr(\tB^{\mu}\tY^A) -i\partial_\mu Y^A_+ \tr(\tB^{\mu}\tY_A^\dag)\nonumber \\
&=&-\partial_\mu Y_{+}^A\partial^\mu Y_{-A}^{\dag}-\partial_\mu Y_{+ A}^\dag
\partial^\mu Y_{-}^A-\tr (\tD_\mu \tY^\dag_A \tD^\mu \tY^A)\nonumber \\
&&+ i\partial_\mu Y_{+A}^\dag\tr(\tB^{\mu}\tY^A) -i\partial_\mu Y^A_+ \tr(\tB^{\mu} \tY_A^\dag)
\end{eqnarray}
In order to derive the kinetic term for the L-BLG theory which only contains real scalar field, we first split the complex scalar
field of the ABJ theory in to two real scalar fields in the following way
\begin{eqnarray}\label{rdf}
&&Y^A_\pm=X^{2A-1}_\pm+iX^{2A}_\pm \ ,\quad  Y^{\dag}_{A\pm}=X^{2A-1}_\pm-iX^{2A}_\pm ,\nonumber \\
&&\tY^A=-\tX^{2A}+i\tX^{2A-1} \ , \quad \tY_A^\dag=-\tX^{2A}-i\tX^{2A-1}.
\end{eqnarray}
Note that the reality of the scalar fields,
$X^A_\pm $ implies $ X^{A*}_\pm=X^A_\pm \ ,
\: \tX^{A\dag}=\tX^A$. In the ABJ theory there are four complex fields denoted by gauge indices $A$ running from 1 to 4.
In our case, we have eight
real scalar fields and we denote them by gauge index $I$ which runs from 1 to 8. Finally using these relations we are able to reproduce the exact
form of the kinetic term for the scalar filed of the L-BLG theory:
\begin{equation}\label{kinlblg}
-2\partial_\mu X^I_+\partial^\mu X^I_--2\partial_\mu X^I_+ \tr (\tB^\mu \tX^I)-\tr[D_\mu \tX^I-\tB_\mu X^I_+]^2,
\end{equation}
where we have defined
\begin{equation}
D_\mu \tX^I=\partial_\mu \tX^I+
i[A_\mu,\tX^I] \ .
\end{equation}

\noindent
We now consider the scaling limit of the kinetic term for fermions. We follow the same procedure as described in the scalar field.
We insert (\ref{scalgauge}), (\ref{scalscalar}) and (\ref{SU}) into the kinetic term for fermion and take the
limit $\epsilon \rightarrow 0$. This finally yields
\begin{eqnarray}\label{kinf}
\hspace{-.5in} -i\tr (\overline{\psi}^{A}\gamma^\mu D_\mu\psi_A)&=& -\frac{i N}{\epsilon^2}
\overline{\psi}^{A}_{+}\gamma^\mu \partial_\mu \psi_{A+} -i\tr (\overline{\theta}^{A}\gamma^\mu \tilde{D}_\mu \theta_A)-
\overline{ \psi}^{A}_+\gamma^\mu\tr(\tB_\mu\theta_A).
 \end{eqnarray}
As in the case of scalar kinetic term, the first term again diverges in the limit $\epsilon \rightarrow 0$. We resolve this issue
 in a similar spirit as in the scalar case, by adding an extra ghost contribution in the ABJ Lagrangian of the form
\begin{eqnarray}\label{ghostf}
i\tr(\overline{V}^{A}
\gamma^\mu\tD_\mu V_A )).
\end{eqnarray}
Here $V^A$ is fermionic field and we redefine this as
\begin{equation}
V_A=-\frac{1}{\epsilon}\psi_{+A}{\rm I}_{N\times N}+\frac{\epsilon}{N}\psi_{-A}{\rm I}_{N\times N}.
\end{equation}
Just like in the scalar case, we introduce a new fermion field
$\psi_{-A}$ through the redefinition of the fermionic ghost field.
Finally, we find that the kinetic term for fermions with
(\ref{ghostf}) taken into account which is now well defined even
in the limit $\epsilon\rightarrow 0$ and it takes the form
\begin{eqnarray}{\label{kfermighost}}
-i\tr(\overline{\theta}^{A}\gamma^\mu \tilde{D}_\mu \theta_A)-\overline{\psi}_+^{A}\gamma^\mu \tr(\tB_\mu\theta_A)
-i\overline{\psi}_+^{A}\gamma^\mu\partial_\mu \psi_{-A}-i\overline{\psi}_-^{A}\gamma^\mu\partial_\mu \psi_{+A}.
\end{eqnarray}
In order to get to the form of the L-BLG theory we now write down the complex fermion field as a combination of two real fermion:
\begin{eqnarray}\label{rdf1}
&&\overline\psi^{A}_\pm=\overline\Psi_{\pm 2A-1}+i\overline\Psi_{\pm 2A} \ ,\quad
\psi_{\pm A}=\Psi_{\pm 2A-1}+i\Psi_{\pm 2A}\ ,\nonumber \\
&&\overline\theta^{A}=-\overline\Psi_{2A}-i\overline\Psi_{2A-1} \ ,\quad\quad
\theta_A=-\Psi_{2A}+i\Psi_{2A-1} \ .
\end{eqnarray}
Using the above relations, the expression of equation (\ref{kfermighost}) can now be rewritten as
\begin{eqnarray}
-i\tr(\overline\Psi_I \gamma^\mu D_\mu \Psi_I)- 2i\overline\Psi_{+I} \gamma^\mu\tr( \tB_\mu \Psi_{I})
-2i\overline\Psi_{-I} \gamma^\mu \partial_\mu \Psi_{+I}.
\end{eqnarray}
This final expression is exactly the $SO(8)$ invariant fermionic kinetic term of L-BLG model.\\

\noindent
Having the analysis on kinetic terms, we now consider the scaling limit in the potential terms in (\ref{defABJL}) and
(\ref{CSact}). In order to take the scaling limit we follow the same approach as in \cite{Honma}. The extra job in our case,
due to our convention, is that we have to explicitly show that the modes $Z^A$ decouple in the scaling limit. For an example,
consider the following term of the bosonic potential (\ref{CSact}),
\begin{equation}
\frac{1}{k^2}\tr (Y^BY_B^\dag Y^CY_C^\dag Y^AY_A^\dag) \ .
\end{equation}
By then using the scaling limit of scalar fields of equation (\ref{scalscalar}), we compute
\begin{eqnarray}
&&\frac{1}{k^2}\tr (Y^BY_B^\dag Y^CY_C^\dag Y^AY_A^\dag)\nonumber \\
&=&\frac{\epsilon^2}{\tilde{k}^2}\tr (\left(\begin{array}{cc} \epsilon^4 Z^BZ_B^\dag & \epsilon
Z^BY_{+B}^\dag {\rm I}_{N\times N}+\epsilon^2Z^B\tY_B^\dag \\
\epsilon Y^B_+Z_B^\dag +\epsilon^2\tY^B Z_B^\dag & \frac{1}{\epsilon^2}
Y_+^B Y^\dag_{+B}{\rm I}_{N\times N}+\frac{1}{\epsilon}(Y_+^B\tY_B^\dag+
\tY^BY_{+B}^\dag )+ \tY^B\tY_B^\dag\\
\end{array}\right)^3\nonumber \\
&=&\frac{\epsilon^2}{\tilde{k}^2} \tr (\frac{1}{\epsilon^2}Y_+^A Y^\dag_{+A}{\rm I}_{N\times N}
+\frac{1}{\epsilon}(Y_+^A\tY_A^\dag+ \tY^AY_{+A}^\dag)+ \tY^A\tY_A^\dag)^3+O(\epsilon^4).
\end{eqnarray}
Note that the modes $Z^A$ really decouple in the limit $\epsilon \rightarrow 0$. Further, the final expression is exactly in the same form as in the potential of
$U(N)\times U(N)$ ABJM theory and diverges as usual in the limit $\epsilon\rightarrow 0$. Again we can resolve this issue following the
analysis of \cite{Honma}. We write the potential as a sum of $V_B^{(n)}$, $V_B=\sum_{n=0}^6 V_B^{(n)}$. Where $V_B^{(n)}$ contains $n$ $Y_+$
fields and $(6-n)$ $\tY$ fields and also $V_B^{(n)}$ scales as $\epsilon^{2-n}$ in the limit $\epsilon\rightarrow 0$.
Then it is obvious that potential terms with $n<2$ vanish in the limit $\epsilon\rightarrow 0$ and to avoid the divergences
the coefficients of the terms, $V_B^{(n)}$ should vanish for $n\geq 3$. Therefore, finally the non-zero contribution comes only from the
$V_B^{(2)}$ part of the potential and these contributing terms combine as
\begin{eqnarray}
&&\frac{4\pi^2}{3\tilde{k}^2}\tr [Y^A_+Y_{+A}^\dag \tY^B\tY_B^\dag\tY^C\tY_C^\dag+Y^B_+Y_{+B}^\dag
\tY^C\tY_C^\dag \tY^A\tY_A^\dag+Y^C_+Y_{+C}^\dag \tY^A\tY_A^\dag \tY^B\tY_B^\dag \nonumber \\
&&+(Y_+^A\tY_A^\dag+\tY^A Y_{+A}^\dag)(Y^B_+\tY_B^\dag+\tY^BY_{+B}^\dag)\tY^C\tY_C^\dag+
 (Y^C_+\tY_C^\dag+\tY^CY_{+C}^\dag) (Y_+^A\tY_A^\dag+\tY^AY_{+A}^\dag)\tY^B\tY_B^\dag\nonumber \\
&&+(Y_+^B\tY_B^\dag+\tY^B Y_{+B}^\dag)(Y^C_+\tY_C^\dag+\tY^CY_{+C}^\dag)\tY^A\tY_A^\dag \nonumber \\
&&+Y_{+A}^\dag Y^A_+ \tY_B^\dag\tY^B\tY_C^\dag\tY^C +Y_{+B}^\dag Y^B_+\tY_C^\dag\tY^C\tY_A^\dag \tY^A
+Y_{+C}^\dag Y^C_+ \tY_A^\dag\tY^A\tY_B^\dag\tY^B\nonumber \\
&&+(Y_{+A}^\dag \tY^A+\tY_A^\dag Y^A_+)(Y_{+B}^\dag \tY^B+\tY_B^\dag Y^B_+)\tY_C^\dag \tY^C+
 (Y_{+C}^\dag\tY^C+\tY_C^\dag Y^C_+) (Y_{+A}^\dag\tY^A+\tY_A^\dag Y^A_+) \tY_B^\dag\tY^B\nonumber \\
&&+(Y_{+B}^\dag \tY^B+\tY_B^\dag Y^B_+)(Y_{+C}^\dag \tY^C+\tY_C^\dag Y^C_+)\tY_A^\dag \tY^A \nonumber \\
&&+4(Y^A_+Y_{+B}^\dag \tY^C\tY_A^\dag\tY^B\tY_C^\dag+ Y^C_+Y_{+A}^\dag\tY^B\tY_C^\dag \tY^A\tY_B^\dag+
Y^B_+Y_{+C}^\dag \tY^A\tY_B^\dag\tY^C\tY_A^\dag\nonumber \\
&&+(Y^A_+\tY_B^\dag+\tY^A Y_{+B}^\dag)(Y^C_+\tY_A^\dag+\tY^CY_{+A}^\dag)\tY^B\tY_C^\dag+
(Y^B_+\tY_C^\dag+\tY^B Y_{+C}^\dag)(Y^A_+\tY_B^\dag+\tY^AY_{+B}^\dag)\tY^C\tY_A^\dag \nonumber\\
&&+(Y^C_+\tY_A^\dag+\tY^C Y_{+A}^\dag)(Y^B_+\tY_C^\dag+\tY^BY_{+C}^\dag)\tY^A\tY_B^\dag \nonumber\\
&&-6[Y^A_+Y_{+B}^\dag \tY^B\tY_A^\dag\tY^C\tY_C^\dag+Y^B_+Y_{+A}^\dag\tY^C\tY_C^\dag \tY^A\tY_B^\dag+
Y_+^C Y_{+C}^\dag \tY^A\tY_B^\dag\tY^B\tY_A^\dag \nonumber \\
&&+(Y_+^A\tY_B^\dag+\tY^A Y_{+B}^\dag)(Y^B_+\tY_A^\dag+\tY^BY_{+A}^\dag)\tY^C\tY_C^\dag+
(Y_+^C\tY_C^\dag+\tY^C Y_{+C}^\dag)(Y^A_+\tY_B^\dag+\tY^AY_{+B}^\dag)\tY^B\tY_A^\dag\nonumber \\
&&+(Y_+^B\tY_A^\dag+\tY^B Y_{+A}^\dag)(Y^C_+\tY_C^\dag+\tY^CY_{+C}^\dag)\tY^A\tY_B^\dag].\nonumber \\
\end{eqnarray}
As earlier, by decomposing the complex scalar fields into two real scalar fields by using  equation (\ref{rdf})
we can get to the form of the L-BLG theory. This analysis has already been done in \cite{Honma} for ABJM.
We are not going to repeat this here but write down the final form of this sextic potential:
\begin{eqnarray}\label{spot}
V_{pot}=\frac{1}{12}\tr (X_+
^I[X^J,X^K]+X^J_+[X^K,X^I]+X^K_+[X^I,X^J])^2.
\end{eqnarray}
In the same way, we can also analyze the potential terms in (\ref{defABJL}) that contain both fermions and bosons. Let us
start with the expression
\begin{equation}\label{FS}
\frac{2\pi}{k} \tr(\overline{\psi}^{A}\psi_A Y_B^\dag Y^B-\psi_B\overline{\psi}^{B}Y^A Y^{\dag}_A ).
\end{equation}
Then using the scaling of
(\ref{scalscalar}) and
$k=\frac{1}{\epsilon}\tilde{k}$ we obtain that in  the limit $\epsilon\rightarrow 0$ (\ref{FS}) reduces to
\begin{eqnarray}
&& \frac{2\pi}{k} \tr(\overline{\psi}^{A}\psi_A Y_B^\dag Y^B-\psi_B\overline{\psi}^{B}Y^A Y^{\dag}_A )\rightarrow\nonumber\\
 &&\frac{2\pi}{\tilde{k}\epsilon^3} \tr(\overline{\psi}_+^{A}{\rm I}_{N\times N}+\epsilon \overline{\theta}^{A})
\times (\psi_{+A}{\rm I}_{N\times N}+\epsilon \theta_A) \times(Y_{+B}^\dag {\rm I}_{N\times N}+ \epsilon\tY_B^\dag)
(Y^B_{+}{\rm I}_{N\times N}+\epsilon\tY^B)\nonumber \\
&&-\frac{2\pi}{\tilde{k}\epsilon^3} \tr(\psi_{+B}{\rm I}_{N\times N}+\epsilon \theta_B)
(\overline{\psi}_+^{B}{\rm I}_{N\times N}+\epsilon\overline{\theta}^{B})
(Y_+^A {\rm I}_{N\times N}+\epsilon\tY^A)(Y_{+A}^\dag {\rm I}_{N\times N}+\epsilon \tY_A^\dag)\nonumber \\
&&=\frac{2\pi}{\tilde{k}\epsilon^3} \tr(\overline{\psi}_+^{A}\psi_{+A} Y^B_+Y_{+B}^\dag {\rm I}_{N\times N}-
\psi_{+B}\overline{\psi}_+^{B}Y^A_+Y_{+A}^\dag {\rm I}_{N\times N})+\frac{2\pi}{\tilde{k}\epsilon^2} \tr(\cdots).
\end{eqnarray}
The term proportional to $\frac{1}{\epsilon^3}$ vanishes due to the fermionic nature of $\psi$. In the
same way we can show that all terms containing $\overline{\psi}_+^{A}\psi_{A+}$ vanish.  Terms proportional to
$\frac{1}{\epsilon^2}$ vanish due to the fact that $\tr(\cdots)$ is zero since it contains either $\theta$ or $\tY$.
The coefficient of the term proportional to $\frac{1}{\epsilon}$ is also zero because this term
contains two traceless fields. Therefore the only non-zero terms are proportional to $\epsilon^0$. In fact,
this analysis is the same as in \cite{Honma} and again we will not repeat the same here. We just write down
the potential term that contains both bosons and fermions. This looks like
\begin{equation}
V_f^{scale} = \overline\Psi_{+L} X^I[X^J,\Gamma_{IJ}\Psi_{L}] - \overline\Psi_{L} X_+^I[X^J,\Gamma_{IJ}\Psi_{L}]
\end{equation}
The $\Gamma$ matrices are the $8\times 8$ matrices and are constructed from the direct product of $\gamma^\mu$.
The form of this gamma matrices is also given in \cite{Honma}.\\

\noindent
After finishing the discussion on undeformed part of the Lagrangian, we now move to the scaling limit
of the mass deformed part of the Lagrangian with the aim of getting contribution for the mass deformed L-BLG theory.
The mass term we consider here is;
\begin{eqnarray}{\label{Vmass}}
V_{mass}&=& m^2 \tr (Y^\dag_AY^A)+im\tr(\overline \psi^{ a}\psi_a)-im\tr(\overline\psi^{a'} \psi_{a'})\nonumber \\
&&-\frac{4\pi}{k}m\tr (Y^aY_{[a}^\dag Y^bY_{b]}^\dag- Y^{a'}Y_{[a'}^\dag Y^{b'}Y^{\dag}_{b']}).
\end{eqnarray}
By performing the scaling limit in the first term we find
\begin{eqnarray}\label{scmassd}
V_{sc.mass}&=& m^2 \tr(Y^\dag_A Y^A)=\frac{m^2}{\epsilon^2}N Y^\dag_{+A}Y^A_++ m^2 \tr (\tY_{A}^\dag \tY^A).
\end{eqnarray}
We see that in the same way as in case of kinetic term here also we need to add the ghost contribution to cancel the divergence
 of the scalar mass term. The ghost contribution is of the form
\begin{equation}
V_{sc.mass}^{ghost}= -m^2 \tr (U_A^\dag U^A)=-\frac{N}{\epsilon^2}m^2 Y_{+A}^\dag
Y^A+m^2 (Y_+^A Y_{-A}^\dag+ Y_{+A}^\dag Y^A_-).
\end{equation}
Finally mass term with the addition of ghost term looks like
\begin{eqnarray}\label{scmass}
V_{sc.mass}=  m^2 \tr (\tY_{A}^\dag \tY^A)+ m^2 (Y_+^A Y_{-A}^\dag+ Y_{+A}^\dag Y^A_-).
\end{eqnarray}
Using the redefinition of the fields introduced in equation ({\ref{rdf}), we find,  $V_{sc.mass}$  can be rewritten
in the form of scalar mass term of L-BLG theory:
\begin{eqnarray}
&& m^2 \tr (\tY_{A}^\dag\tY^A)+ m^2 (Y_+^A Y_{-A}^\dag+Y_{+A}^\dag Y^A_-)\nonumber \\
&=&2m^2 (X_+^{2A-1}X_-^{2A-1}+X_{+}^{2A}X_-^{2A})+m^2\tr(\tX^{2A}\tX^{2A}+\tX^{2A-1}\tX^{2A-1})\nonumber \\
&=&2m^2 (X_+^IX_-^I)+m^2\tr(\tX^I\tX^I).
\end{eqnarray}
In case of the fermion mass term we find
\begin{eqnarray}
i m \tr(\overline{\psi}^{a}\psi_a)= \frac{iN}{\epsilon^2}m\overline{\psi}^{a}_+ \psi_{+a}+ im\tr (\overline{\theta}^{a}\theta_a)
\end{eqnarray}
and the same for $\psi_{a'}$. Clearly again we have to add ghost contribution to the Lagrangian
to get a finite contribution. The form of this ghost is
\begin{eqnarray}
\mL_{m.g.f.}&=&-im \tr\overline{V}^{a}V_a+im\overline{V}^{a'}V_{a'}\nonumber \\
&=&-im\frac{N}{\epsilon^2}\overline{\psi}^{a}_+\psi_{+a}+im(\overline{\psi}^{a}_+
\psi_{-a}+\overline{\psi}^{a}_-\psi_{+a})\nonumber\\
&&+im\frac{N}{\epsilon^2}\overline{\psi}^{a'}_+\psi_{+a'}-im(\overline{\psi}^{a'}_+\psi_{-a'}+
\overline{\psi}^{a'}_-\psi_{+a'}).
\end{eqnarray}
Clearly, terms proportional to $\frac{1}{\epsilon^2}$ cancel and we end up with the finite result
\begin{eqnarray}
\hspace{-.3in}V_{f.mass}= im \tr(\overline{\theta}^{a}\theta_{a}-\overline{\theta}^{a'}\theta_{a'})
+im(\overline{\psi}^{a}_+\psi_{-a}+\overline{\psi}^{a}_-\psi_{+a}) -im(\overline{\psi}^{a'}_+\psi_{-a'}+
\overline{\psi}^{a'}_-\psi_{+a'}).
\end{eqnarray}
As in the previous cases, we would again redefine the $\psi$ and $\theta$ fields as in equation (\ref{rdf1}) and finally
 we are able to write down $V_{f.mass}$ as the mass term of L-BLG theory.
\begin{eqnarray}
V_{f.mass}&=&im\tr(\overline\Psi_A\Psi_A-\overline\Psi_{A'}\Psi_{A'})
+ 2im(\overline\Psi_{+A}\Psi_{-A}
-\overline\Psi_{+A'}\Psi_{-A'}).
\end{eqnarray}
Here $A$ runs from 1 to 4 and $A'$ runs from 5 to 8.\\

\noindent
As the next step we analyze the first term of mass deformed potential with the following
expression
\begin{eqnarray}
V_{d.pot1}&=&-\frac{4\pi}{k} \tr (Y^a Y_{[a}^\dag Y^b Y_{b]}^\dag)\nonumber\\
 &=&-\frac{4\pi}{\tilde{k}}[\frac{N}{\epsilon^3}Y^a_+Y_{+[a}^\dag Y^b_+Y_{+b]}^\dag
+\frac{1}{\epsilon} \tr(Y^a_+Y_{+[a}^\dag\tY^b\tY^\dag_{b]}+ \tY^a\tY^\dag_{[a}
Y^b_+Y_{+b]}^\dag \nonumber\\
&&+(Y^a_{+}\tY^\dag_{[a}+\tY^a Y_{+[a}^\dag)(Y^b_{+}\tY^\dag_{b]}+
 \tY^b Y_{+b]}^\dag))\nonumber\\
&&+\tr(( Y^a_{+}\tY^\dag_{[a}+\tY^a Y_{+[a}^\dag)\tY^b\tY^\dag_{b]}+
 \tY^a \tY^\dag_{[a}( Y^b_{+}\tY^\dag_{b]}+ \tY^b Y_{+b]}^\dag))].  \nonumber \\
\end{eqnarray}
Now the first term vanishes:
\begin{eqnarray}
Y^a_+Y_{+[a}^\dag Y^b_+Y_{+b]}^\dag&=&Y^a_+Y_{+a}^\dag Y^b_+Y_{+b}^\dag
-Y^a_+Y_{+b}^\dag Y^b_+Y_{+a}^\dag\nonumber \\
&=&Y^a_+Y_{+a}^\dag Y^b_+Y_{+b}^\dag-Y^a_+Y_{+a}^\dag Y^b_+ Y_{+b}^\dag =0.
\end{eqnarray}
Let us now analyze the contributions proportional to $\frac{1}{\epsilon}$
\begin{eqnarray}
 &&\tr(Y^a_+Y_{+[a}^\dag\tY^b\tY^\dag_{b]}+ \tY^a\tY^\dag_{[a}Y^b_+Y_{+b]}^\dag
+(Y^a_{+}\tY^\dag_{[a}+ \tY^a Y_{+[a}^\dag)(Y^b_{+}\tY^\dag_{b]}+ \tY^b Y_{+b]}^\dag) \nonumber\\
&=&\tr(2  Y^a_+Y_{+[a}^\dag\tY^b\tY^\dag_{b]}-Y^a_+Y^\dag_{+a}\tY^b\tY_{b}^\dag+
Y^a_+Y^\dag_{+b}\tY^b\tY_a^\dag-(Y^a_+Y^\dag_{+a}\tY^b \tY_b^\dag-
Y^a_+Y^\dag_{+b}\tY^b\tY_a^\dag))\nonumber \\
&=&\tr(2  Y^a_+Y_{+[a}^\dag\tY^b\tY^\dag_{b]} -2Y^a_+Y_{+[a}^\dag
\tY^b\tY^\dag_{b]})=0\nonumber \\
\end{eqnarray}
using
\begin{eqnarray}
Y^a_{+}\tY^\dag_{[a}Y^b_{+}\tY^\dag_{b]}=Y^a_+\tY_a^\dag Y_+^b
\tY_b^\dag-Y^a_+ \tY^\dag_bY^b_+\tY^\dag_a=Y^a_+Y^b_+\tY_a^\dag \tY_b^\dag-
Y^a_+Y^b_+\tY_a^\dag \tY_b^\dag=0.\nonumber \\
\end{eqnarray}
Then we find the the finite contribution to the potential;
\begin{eqnarray}\label{vmp}
V_{d.pot1}&=&-\frac{4\pi}{\tilde{k}} m\tr(Y^a_+\tY^\dag_a \tY^b\tY^\dag_b-
Y^a_+\tY^\dag_b\tY^b\tY^\dag_a+\tY^a Y^\dag_{+a}\tY^b \tY^\dag_b
-\tY^a Y^\dag_{+b}\tY^b \tY^\dag_a)\nonumber \\
&&+\tY^a\tY_a^\dag Y^b_+\tY^\dag_b-\tY^a\tY_b^\dag
Y_+^b\tY_a^\dag+\tY^a\tY_a^\dag \tY^b Y_{+b}^\dag-
\tY^a\tY_b^\dag \tY^b Y_{+a}^\dag)\nonumber \\
 &=&-\frac{4\pi}{\tilde{k}}m\tr (Y^a_+\tY_b^\dag(\tY_a^\dag
\tY^b-\tY^b \tY_a^\dag)+Y^\dag_{+a}\tY^\dag_b(\tY^a\tY^b-\tY^b\tY^a)\nonumber \\
&&+Y^b_+\tY^a(\tY^\dag_a\tY^\dag_b-\tY^\dag_b\tY^\dag_a)+ Y^\dag_{+a}\tY^b
(\tY^\dag_b\tY^a-\tY^a\tY_b^\dag)).
\end{eqnarray}
We now again redefine the field $Y'$s in the previous way as in
equation (\ref{rdf}) and with these redefinition we can write the
above expression as
\begin{eqnarray}
&&-4iX^{2a-1}_+\tX^{2b}[\tX^{2b-1},\tX^{2a}] -4i X^{2a}_+\tX^{2b-1}[\tX^{2b},\tX^{2a-1}]\nonumber \\
&&-4iX^{2a-1}_+\tX^{2b-1}[\tX^{2a},\tX^{2b}]-4iX_+^{2a}\tX^{2b}[\tX^{2a-1},\tX^{2b-1}].
\end{eqnarray}
Let us now analyze these expressions. The expressions on the first term of the first line give
\begin{eqnarray}
&&-4i X^{2a-1}_+\tr(\tX^{2b}[\tX^{2b-1},\tX^{2a}])\nonumber \\
 {\mathrm for}\, a=b:\quad &&-4i\tr(\tX^{2b-1}\tX^{2b}\tX^{2b}
-\tX^{2b-1}\tX^{2b}\tX^{2b})=0 \ ,\nonumber \\
\mathrm{for}\, a=1,b=2: \quad &&-4iX^{1}_+\tr(\tX^{4} [\tX^{3},\tX^{2}] )
\ , \nonumber \\
\mathrm{for} \, a=2,b=1: \quad && -4i X^3_+\tr(\tX^2[\tX^1,\tX^4])
\end{eqnarray}
and the expression on the last term of the first line gives
\begin{eqnarray}
&&-4i X^{2a}_+\tr( \tX^{2b-1}[\tX^{2b},\tX^{2a-1}])\nonumber\\
\mathrm{for} \, a=b: \quad && 0 \nonumber \\
\mathrm{for} \, a=1 \ , b=2: \quad && \ -4i
X^{2}_+\tr( \tX^{3} [\tX^{4},\tX^{1}])\ ,\nonumber \\
\mathrm{for} \, a=2 \ , b=1: \quad && -4iX^4_+\tr (\tX^1[\tX^2,\tX^3]).
\end{eqnarray}
On the other hand the expressions on the second line give
\begin{eqnarray}
&&-4iX^{2a-1}_+\tr (\tX^{2b-1}[\tX^{2a},\tX^{2b}])-4i
X_+^{2a}\tr(\tX^{2b}[\tX^{2a-1},\tX^{2b-1}])\nonumber \\
&&\mathrm{for} \ a=b : \quad  0\nonumber \\
&&\mathrm{for} \quad a=1 , \ b=2:-4iX^{1}_+\tr (\tX^{3}
[\tX^{2},\tX^{4}])-4iX_+^{2}\tr(\tX^{4}[\tX^{1},\tX^{3}])\nonumber \\
&&\mathrm{for} \quad a=2, \ b=1:-4iX^{3}_+\tr (\tX^{1}[\tX^{4},\tX^{2}])-4i
X_+^{4}\tr(\tX^{2}[\tX^{3},\tX^{1}]).
\end{eqnarray}
All together we have
\begin{eqnarray}
&& -8iX^{1}_+\tr (\tX^{3}
[\tX^{2},\tX^{4}])-8iX_+^{2}\tr(\tX^{4}[\tX^{1},\tX^{3}])\nonumber \\
&&-8iX^{3}_+\tr (\tX^{1}[\tX^{4},\tX^{2}])-8i
X_+^{4}\tr(\tX^{2}[\tX^{3},\tX^{1}])
\end{eqnarray}
We can write
\begin{eqnarray}
-6iX^{1}_+\tr (\tX^{3}[\tX^{2},\tX^{4}])&=&i\epsilon_{1324}X^{1}_+\tr (\tX^{3}
[\tX^{2},\tX^{4}])+ i\epsilon_{1342}X^1_+\tr (\tX^3[\tX^4,\tX^2])\nonumber\\
&&+i\epsilon_{1243}X^1_+ \tr (\tX^2[\tX^4,\tX^3])+ i\epsilon_{1234} X^1_+
\tr (\tX^2[\tX^3,\tX^4]). \nonumber \\
&&+i\epsilon_{1423} X^1_+
\tr (\tX^4[\tX^2,\tX^3])+i\epsilon_{1432} X^1_+
\tr (\tX^4[\tX^3,\tX^2])\nonumber \\
\end{eqnarray}
In the same way we can proceed for expressions with primed quantities and in summary, we obtain the scaling limit of the
deformed potential in the form
\begin{equation}\label{dmasspot}
V_{d.pot}=-i\frac{16\pi}{3\tilde{k}}m\epsilon_{ABCD}\tr (X^A_+ \tX^B [\tX^C,\tX^D])
+i\frac{16 \pi}{3\tilde{k}}m\epsilon_{A'B'C'D'} \tr (X^{A'}_+\tX^{B'} [\tX^{C'},\tX^{D'}])
\end{equation}
Finally we would like to sum up all the terms of the mass deformed ABJ Lagrangian with the scaling limit defined and
the sum gives us:
\begin{eqnarray}
\mL_{BLG}&=&-2\partial_\mu X^I_+\partial^\mu X^I_-
-2\partial_\mu X^I_+ \tr (B^\mu X^I)
-\tr[D_\mu X^I-B_\mu
X_+^I]^2\nonumber\\
&&-i\tr(\overline\Psi_I \gamma^\mu D_\mu \Psi_I)-2i\overline\Psi_{+I} \gamma^\mu\tr(B_\mu \Psi_{I})
-2i\overline\Psi_{-I} \gamma^\mu \partial_\mu \Psi_{+I}\nonumber\\
&&-\frac{1}{4}
\epsilon^{\mu\nu\lambda}\tr (B_\mu
F_{\nu\lambda})
+\frac{1}{12} (X_+
^I[X^J,X^K]+X^J_+[X^K,X^I]+X^K_+[X^I,X^J])^2\nonumber\\
&&+ \overline\Psi_{+L} X^I[X^J,\Gamma_{IJ}\Psi_{L}] - \overline\Psi_{L} X_+^I[X^J,\Gamma_{IJ}\Psi_{L}]\nonumber\\
&&+2m^2 (X_+^IX_-^I)+m^2\tr(X^I X^I)
+im\tr(\overline\Psi_A\Psi_A-\overline\Psi_{A'}\Psi_{A'})\nonumber\\
&&+2im(\overline\Psi_{+A}\Psi_{-A}
-\overline\Psi_{+A'}\Psi_{-A'})
- \frac{16}{3} i  m\epsilon_{ABCD}
\tr (X^A_+ X^B [X^C,X^D])\nonumber\\
&&+\frac{16}{3} i m
\epsilon_{A'B'C'D'} \tr (X^{A'}_+
X^{B'} [X^{C'},X^{D'}]),
\end{eqnarray}
where we have done away with the tilde over the fields for notational simplicity. So starting from a mass deformed ABJ theory,
we finally arrived at the  world volume theory of a stack of $N$ number of M2-branes by taking the proper scaling limit on the
fields and simultaneously sending Chern-Simons level, $k$ at infinity. These $N$ number of branes are sitting
far away from the $\mathbf{C^4/Z}_k$ singularity. The final expression of the Lagrangian matches exactly with the mass
deformed L-BLG theory formulated based on $3$-algebra.
%%%%%%%%%%%%%%%%%%%%%%%%%%%%%%%%%%%%%%%%%%%%%%%%%%%%%%%%%%%%%%%%

\section{Conclusion}
In this paper we have started with the undeformed ABJ theory and
then added suitable mass terms which preserve the supersymmetry of
the theory. Our motivation to add those mass terms was to obtain
the mass deformed L-BLG theory. In our construction we have
introduced two new parameters, $\epsilon$ and  $\tilde k$. These
two parameters are related to the Chern-Simons term when $\tilde
k= \epsilon k$, where we kept $\tilde k$ at fixed value $\pi$ by
taking $\epsilon\rightarrow 0$ at the end of the computation and
at the same time sending $k$ to infinity. We have further dropped
some of the fields of the ABJ theory due the fact that they are of
the order $\epsilon \rightarrow 0$ . On top of these two new
parameters, we also have added an extra ghost term of $U(1)$
multiplet in the ABJ theory. Eventhough the ghost term decoupled
at the level of ABJ theory, it is effectively coupled at the level
of L-BLG theory by the redefinitions of the fields. Finally we
have successfully constructed the mass deformed L-BLG theory. One
can extend the present analysis to write down the full
non-relativistic ABJ theory and also for L-BLG theory. Further one
can take the same scaling limit, as we have taken to produce the
mass deformed L-BLG theory, on non-relativistic ABJ theory and
check whether this theory also reproduces the non-relativistic
L-BLG theory.
We wish to return to this problem in future.\\

\noindent
{\bf Acknowledgements:} We would like to thank Josef Kluson for participation at
an early stage of this work. It is a pleasure to thank Souvik Benerjee, Shankhadeep Chakrabortty, Sudipta Paul Choudhury, Sumit Das and Sudipta Mukherji
for the fruitful discussions. T.K.D would like to thank Institute of Physics, Bhubaneswar for the hospitality where part of this work was done.
%\newpage
%%%%%%%%%%%%%%%%%%%%%%%%%%%%%%%%%%%%%%
%%%%%%% Thebibligraphy %%%%%%%%%%%%%%%%%%%%%
%%%%%%%%%%%%%%%%%%%%%%%%%%%%%%%%%%%%%

\end{document}